\documentclass{article}

\usepackage{amsmath,amsfonts}

\usepackage[caption=false,font=normalsize,labelfont=sf,textfont=sf]{subfig}
\usepackage{textcomp}
\usepackage{stfloats}
\usepackage{url}
\usepackage{bm}

\usepackage{graphicx}
\usepackage{cite}
\usepackage{siunitx}

\usepackage{tabularx}
\usepackage{booktabs}
\usepackage{authblk}
\usepackage[a4paper, total={6in, 8in}]{geometry}

\title{Frequency- and Power-Dependent Optical Response of a Cesium Rydberg Microwave Receiver}

\author[1,2]{Jure~Pirman}
\author[1,2]{Ema~Stopar}
\author[1,2]{Katja~Gosar}
\author[1,3]{Erik~Zupani\v{c}}
\author[1]{Peter~Jegli\v{c}}
\affil[1]{Jo\v{z}ef Stefan Institute, Jamova 39, SI-1000 Ljubljana, Slovenia.}%
\affil[2]{Faculty of Mathematics and Physics, University of Ljubljana,
Jadranska 19, SI-1000 Ljubljana, Slovenia.}%
\affil[3]{Atom Quantum Labs, Ajdov\v{s}\v{c}ina 1,
SI-1000 Ljubljana, Slovenia.}%

\begin{document}
\maketitle
\begin{abstract}
We show experimentally obtained frequency- and power-dependent optical
response of a room-temperature cesium vapor cell used as a Rydberg
microwave receiver based on electromagnetically induced transparency. An
applied microwave field couples the ${70S_{1/2}}$ Rydberg state to the nearby
${70P_{1/2}}$, ${70P_{3/2}}$,
${69P_{1/2}}$, and ${69P_{3/2}}$ states, modifying the
probe-transmission spectrum. Probe transmission is  recorded over a 800 MHz range of coupling laser detunings. These EIT spectra are measured for RF frequencies from 9 to 13.5 GHz and over a 40 dB range of RF power. We analyze the RF induced splittings in the recorded EIT spectra. We observe both limiting cases, where the EIT peak separation is either dominated by the RF detuning or by the RF power. At higher powers we are also able to observe complex features due to coupling of multiple nearby Rydberg states. We further analyze the probe-transmission signal at zero coupling-laser detuning. Sigmoidal fits to the power-response curves are used to determine the midpoint power and the 10--90\% transition interval as functions of RF frequency.
\end{abstract}

\section{Introduction}
Rydberg-atom vapor cells enable optical detection of radio-frequency (RF) electric fields using electromagnetically induced transparency (EIT)
\cite{Sedlacek2012MicrowaveElectrometry,Yuan2023QuantumSensing,Zhang2024HybridReview,Somaweera2025PhotonicsReview}. In an EIT scheme, a probe and a coupling laser establish a narrow transparency feature in the transmission spectrum of
the probe beam \cite{Zhao2009CsEIT,PopulationRepumping}. RF fields can couple different Rydberg states, which causes changes in the measured EIT spectrum
\cite{Holloway2014BroadbandProbe,Anderson2019SelfCalibratingRFMS,FiberCoupled2018AO}. For an RF field near resonance with a selected transition, sufficiently strong coupling results in an Autler--Townes doublet
\cite{Holloway2014BroadbandProbe,APL2018PowerStandard,FiberCoupled2018AO}. These RF-induced spectral
changes provide an optical measure of the response of the atomic system to
the applied RF frequency and power.

Rydberg-EIT systems have consequently been investigated for self-calibrated RF electric-field metrology \cite{Sedlacek2012MicrowaveElectrometry,Anderson2019SelfCalibratingRFMS,APL2018PowerStandard,IEEEOverview2022}, weak-field detection
\cite{Anderson2020AtomicReceivers,Gordon2019SubHzMixer,Jing2020Superhet,MWAmplitudeMod2024}, as well as atomic receivers and spectrum
analyzers \cite{Cox2018QuantumLimited,AMFMReceiver2020,Meyer2021SpectrumAnalyzer,RydbergReceiver2025IEEE}. Their potential for
wireless communication has been demonstrated through multiband reception \cite{Multiband2022AIP}, video streaming \cite{TVStreaming2022}, and image transmission
\cite{Zhang2024ImageTransmissionRydbergAntenna}. Passive Rydberg-atomic
transducers have also enabled remote RF-field sensing
\cite{Otto2023DistantRFSensingAPL}. These applications require a detailed understanding of how the optical response depends on both the incident RF-field parameters and the optical parameters used to probe the atomic spectrum \cite{Meyer2019AssessmentWideband,ContinuouslyTunable,Miller2016RFModulation,ContinuousBroadband}.

In this work, we experimentally investigate the optical response of a room-temperature cesium vapor cell operated as a Rydberg microwave receiver as a function of RF frequency, RF power, and coupling-laser detuning. First, we present the experimental setup and the relevant optical and RF transitions, together with the two-level model describing the dependence of the spectral splitting on RF power and detuning from a selected Rydberg--Rydberg transition. Then, we extract the separation between the split peaks as a function of RF detuning for applied powers spanning $40~\mathrm{dB}$ and compare the results with the prediction of a two-level model. Next, we extend the spectral measurements over the full investigated RF-frequency range from $9~\mathrm{GHz}$ to $13.5~\mathrm{GHz}$ and present the photodetector signal as a function of coupling-laser detuning and RF frequency. Finally, we examine the optical response with zero coupling field detuning across the full RF-frequency and power range.
Together, these measurements provide a detailed picture of the receivers response over a broad range of RF detunings and powers.

\section{Experimental Setup and Measurement Protocol}
\label{sec:setup}

\begin{figure}[h]
  \centering
  \includegraphics{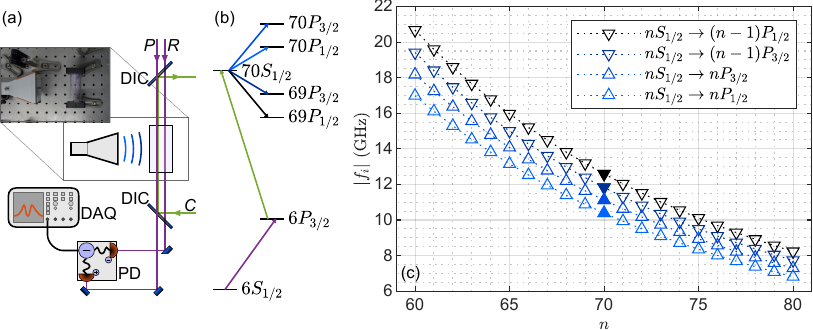}
  \caption{(a) Schematic of the experiment. Probe beam (\textit{P}) and reference beam (\textit{R}) propagate through the cesium vapor cell. The probe beam is overlayed with the counter-propagating coupling (\textit{c}) beam. Probe and coupling beams are combined and separated using dichroic mirrors (DIC). An RF horn antenna is used to generate microwave fields. The probe and reference beams are separated using D-shaped mirror and is guided onto the differential photodetector (PD), where the reference signal is subtracted from the probe. The electric signal is then acquired using a digital acquisition board (DAQ). (b) Energy level diagram used throughout the experiment. (c) Transition frequencies for different Rydberg-Rydber transitions.}
  \label{fig:setup_levels}
\end{figure}

As shown in Fig.~\ref{fig:setup_levels}(a), a weak probe field at \SI{852}{\nano\meter} drives the $6\mathrm{S}_{1/2}\rightarrow6\mathrm{P}_{3/2}$ transition, while a coupling
field at \SI{509}{\nano\meter} drives the $6\mathrm{P}_{3/2}\rightarrow70\mathrm{S}_{1/2}$ transition to the target Rydberg state. This state is then coupled to nearby Rydberg states using microwave fields. The probe-laser is locked to the $F=4\to F'=5$ transition using modulation transfer spectroscopy. The coupling laser is referenced to a high-finesse cavity using the Pound--Drever--Hall technique. The narrow cavity resonances and modulation sidebands provide an accurate frequency ruler used to determine the coupling-laser detuning $\Delta_c$. The probe and coupling beams counter-propagate through a \SI{75}{\milli\meter} long, room-temperature cesium vapor cell without buffer gasses or wall coatings. The experiment is performed without magnetic-field shielding or active magnetic-field compensation. No additional electric-field shielding is used, although the vapor cell walls themselves can significantly attenuate external low-frequency electric fields \cite{Ma2022ElectricFields}. The probe transmission is measured using a differential photodetector with an additional reference beam that also propagates through the vapor cell, but without the coupling beam. Subtracting the reference signal increases the visibility of the small EIT-induced modulation and also suppresses effects such as laser-intensity fluctuations and small changes in the optical density inside the vapor cell. A continuous-wave RF field is applied using a horn antenna (RF SPIN DRH30) with the apperture centered and 75 mm away from the vapor cell. The RF field couples the optically excited Rydberg state $70\mathrm{S}_{1/2}$ to different Rydberg states, namely $70\mathrm{P}_{1/2}$, $70\mathrm{P}_{3/2}$, $69\mathrm{P}_{1/2}$ and $69\mathrm{P}_{3/2}$ as shown in Fig.~\ref{fig:setup_levels}(b). Choosing a different Rydberg state allows coupling to different states with different transition frequencies as shown in Fig.~\ref{fig:setup_levels}(c).

For each RF frequency $f_{\mathrm{RF}}$ and applied RF generator power $P_{\mathrm{RF}}$, we record the complete EIT spectrum as a function of $\Delta_c$. The result of the experiment is a three-dimensional dataset $S(\Delta_c,f_{\mathrm{RF}},P_{\mathrm{RF}})$, from which the spectra and response maps presented below are obtained.

\section{RF-Dressed-State Model}
\label{sec:model}

To interpret the RF-induced splitting observed in the EIT spectra, we treat
the microwave coupling between the $70S_{1/2}$ state and each of the nearby
$69P_{1/2}$, $69P_{3/2}$, $70P_{1/2}$, and $70P_{3/2}$ states as a driven
two-level system. For a given transition, we define the RF detuning as
\begin{equation}
\delta_{\mathrm{RF},i}=f_{\mathrm{RF}}-f_0,
\label{eq}
\end{equation}
where $f_0$ is the corresponding Rydberg--Rydberg transition frequency. The
transition frequencies used throughout this work were calculated using the
Alkali Rydberg Calculator (ARC) \cite{arc} and are listed in
Table~\ref{tab:transitions}, where the sign indicates whether the coupled state lies above or below the $70S_{1/2}$ state.

\begin{table}
\centering
\caption{Calculated and experimentally obtained transition frequencies from the $70S_{1/2}$ state to
nearby $P$ Rydberg states. The positive or negative sign of ${f_i}$ denotes whether the transition is to a higher or lower energy state, respectively. Transition frequencies $f_e$ are obtained by fitting the measured splitting in the EIT signal at different RF powers. Values $|f_i|-|f_e|$ are the difference between the theoretical and experimentally obtained values. The last two entries list the transition frequencies of the degenerate two photon transitions that were observed in the experiment.\label{tab:transitions}}

\begin{tabular}{lccc}

\toprule
Transition	& ${f_0}$ (GHz) & ${f_e}$ (GHz) &$|f_i|-|f_e|$ (MHz)\\
\midrule
$70S_{1/2}\rightarrow70P_{1/2}$ & $+10.39113$ & $10.392\pm0.001$ & $-1$ \\
$70S_{1/2}\rightarrow70P_{3/2}$ & $+11.12351$ & $11.119\pm0.001$ & $+4$\\
$70S_{1/2}\rightarrow69P_{3/2}$ & $-11.82681$ & $11.833\pm0.001$ & $-6$\\
$70S_{1/2}\rightarrow69P_{1/2}$ & $-12.59328$ & $12.593\pm0.001$ & $0$\\
$70S_{1/2}\rightarrow69S_{1/2}$ & $-11.73502\times2$ &-&-\\
$70S_{1/2}\rightarrow71S_{1/2}$ & $+11.21313\times2$ &-&-\\
\bottomrule
\end{tabular}

\end{table}

Within this model, the square of frequency separation between the
two RF-coupled resonances is \cite{Zhang2019DetuningRF}
\begin{equation}
    \Delta_f^2
    =
        \delta_{\mathrm{RF}}^2
        +
        \left(
            \frac{\Omega_{\mathrm{RF}}}{2\pi}
        \right)^2,
    \label{eq:generalized_splitting}
\end{equation}
\noindent
where $\Omega_{\mathrm{RF}}$ is the RF Rabi frequency. 
\noindent
Eq.~\eqref{eq:generalized_splitting} has two limiting regimes. For low detuning regime, $|\delta_{\mathrm{RF}}| \ll \Omega_{\mathrm{RF}}/(2\pi)$,
the splitting is dominated by the RF coupling,
\begin{equation}
\Delta_f \approx \frac{\Omega_{\mathrm{RF}}}{2\pi},
\label{eq:coupling}
\end{equation}
while in the high detuning regime $|\delta_{\mathrm{RF}}| \gg \Omega_{\mathrm{RF}}/(2\pi)$,
it is dominated by the detuning,
\begin{equation}
\Delta_f \approx |\delta_{\mathrm{RF}}|=|f_{\mathrm{RF}}-f_0|.
\label{eq:detuning}
\end{equation}

\section{Results and Discussion}
\label{sec:results}

\subsection{Power-Dependent EIT Spectra}
\label{subsec:stacked}
\begin{figure}[h]
  \centering
  \includegraphics{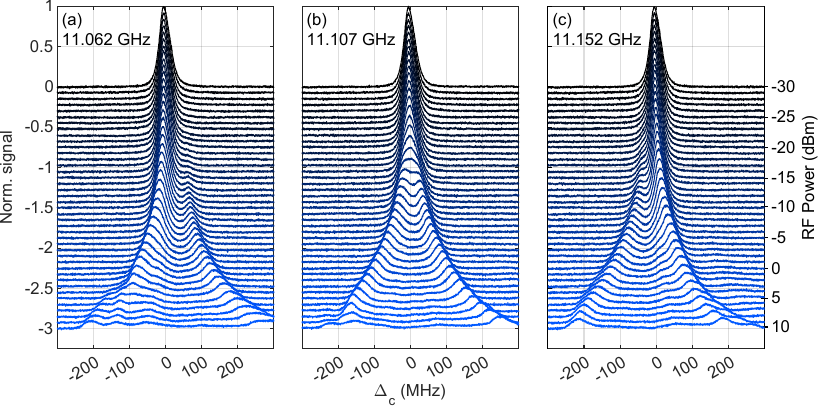}
  \caption{Normalized EIT signals for different RF powers, at RF frequencies (a) 11.062 GHz, (b) 11.107 GHz, and (c) 11.152 GHz. These correspond to detunings $\delta_{\mathrm{RF}}$ of $-57$, $-12$, and $+33$ MHz from the experimentally determined transition frequency $f_0 = 11.119\ \mathrm{GHz}$ (theoretical value $f_0 = 11.124\ \mathrm{GHz}$, per Table (\ref{tab:transitions}). }
  \label{fig:fig1}
\end{figure}

We first examine how the EIT spectrum depends on the applied RF power in proximity to the selected $70\mathrm{S}_{1/2}\rightarrow 70\mathrm{P}_{3/2}$ transition. Fig.~\ref{fig:fig1} shows spectra recorded below, near, and above the transition frequency. 

At the lowest RF powers, no resolved splitting is observed. As the applied power increases, the frequency difference between the peaks $\Delta_f$ increases. The spectra recorded farther from resonance exhibit asymmetrically split peaks, whereas the near-resonant spectra show a nearly symmetric splitting.

\subsection{Detuning-Dependent Peak Splitting}
\label{subsec:splitting}

To quantify the RF-induced spectral splitting, each spectrum exhibiting two resolvable peaks is fitted with a sum of two Voigt profiles. We find that pure Gaussian or Lorentzian fits give poorer results, since both Doppler and homogeneous broadening contribute to the measured linewidth. The frequency difference $\Delta_f$ between the split peaks is obtained from the separation between the fitted peak centers. At high RF powers, additional peaks can appear in the spectra, leading to larger fitting uncertainties.

\begin{figure}[h]
  \centering
  \includegraphics{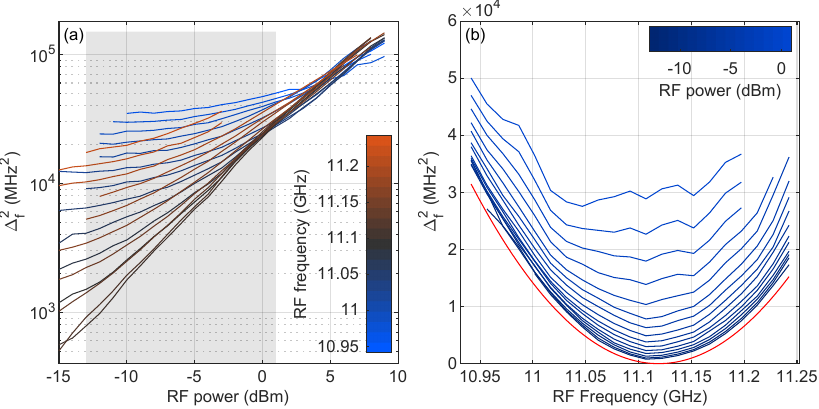}
  \caption{Square of frequency difference $\Delta_f^2$ between largest peaks observed in the EIT signal for different RF powers and frequencies. Frequencies are chosen near the $70\mathrm{S}_{1/2}\to70\mathrm{P}_{3/2}$ transition. (a) depicts the dependence on the RF power. At high RF powers or low detunings from resonance $\Delta_f$ has a power law scaling. (b) shows the dependence of splitting $\Delta_f$ as a function of detuning from resonance. Red line shows the limit value where the RF power approaches 0. Increasing the RF power shifts the $\Delta_f^2$ parabola to higher values without modifying the general shape.}
  \label{fig:fig2}
\end{figure}

Fig.~\ref{fig:fig2}(a) shows $\Delta_f^{2}$ as a function of applied RF power for several RF frequencies. At small detunings, the curves closely follow the resonant power dependence obtained from Eq.~\eqref{eq:coupling}. As the RF frequency is tuned away from the transition, the detuning term $\delta_{\mathrm{RF}}^{2}$ introduces an
offset that is most visible at low powers. At higher powers, the RF Rabi-frequency contribution dominates and the curves approach a common high-power dependence corresponding to the resonant case, $\delta_{\mathrm{RF}}=0$.

Fig.~\ref{fig:fig2}(b) shows $\Delta_f^{2}$ as a function of RF frequency for several applied powers. Each curve has a minimum near the transition frequency and increases with RF detuning. In the weak-field limit, the Rabi-frequency contribution is negligible and Eq.~\eqref{eq:generalized_splitting} reduces to $\Delta_f^{2}=\delta_{\mathrm{RF}}^{2}$, shown by the red parabola. Increasing the applied RF power adds the term $(\Omega_{\mathrm{RF}}/2\pi)^2$, shifting the parabolic dependence toward larger values of $\Delta_f^{2}$ without changing its general shape. A fit to the experimental data gives a transition frequency of $f_0=(11.119\pm0.001) \, \mathrm{GHz}$ for the $70S_{1/2}\rightarrow70P_{3/2}$ transition, consistent with the calculated value of $11.12351 \, \mathrm{GHz}$. Similarly, we extract the transition frequencies for all observed transitions and present them in Table \ref{tab:transitions}.

\subsection{Frequency-Resolved Spectral Maps}
\label{subsec:maps}

\begin{figure}[h]
        \centering
        \makebox[0pt]{\includegraphics{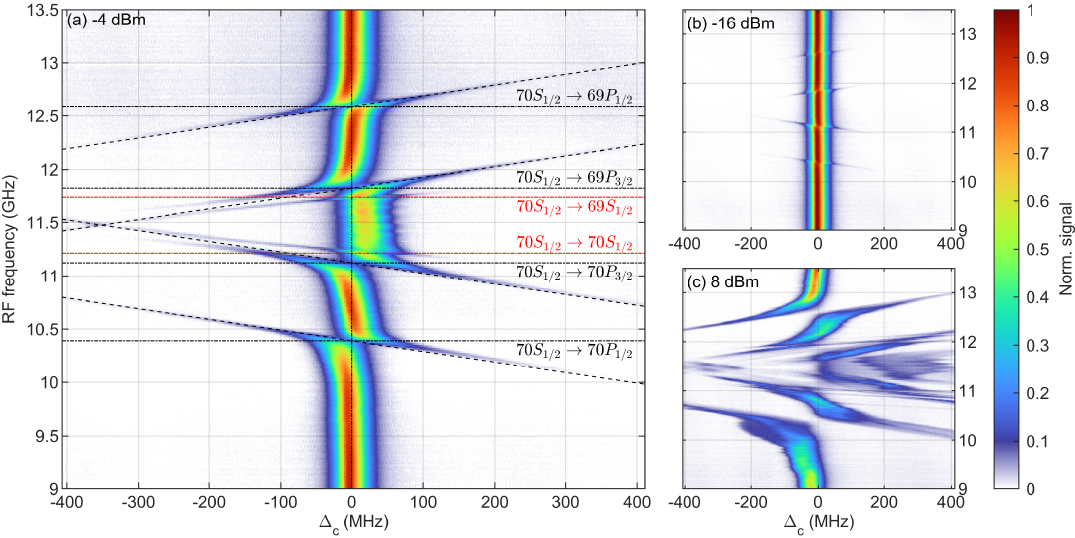}}
        \caption{False color plot of normalized EIT signals for different RF frequencies and powers. Transitions to 4 different states are visible with their respective theoretical frequencies marked with horizontal dash-dotted lines. Dashed lines depict frequencies $f = f_{i}\pm\Delta_c$, and should serve as an approximate position of a second peak in the EIT signal at high values of detuning $\Delta_c$. The red dash-dotted lines indicate the theoretical frequencies of degenerate two-photon transitions.}
        \label{fig:fig3}
\end{figure}

To obtain a broader view of the frequency-dependent response, we scan the RF frequency from $9~\mathrm{GHz}$ to $13.5~\mathrm{GHz}$ and record the complete EIT spectrum at each frequency. Fig.~\ref{fig:fig3} shows the resulting optical signal as a function of coupling-laser detuning and RF frequency for three representative RF powers: $-4~\mathrm{dBm}$ in (a), $-16~\mathrm{dBm}$ in (b), and $8~\mathrm{dBm}$ in (c). 

Fig.~\ref{fig:fig3}(a) clearly shows spectral features associated
with four nearby Rydberg--Rydberg transitions. Their calculated frequencies
are marked by horizontal dash-dotted lines and are listed in
Table~\ref{tab:transitions}. The position of the smaller peak follows the large-detuning behavior predicted by Eq.~\eqref{eq:detuning}. The direction or sign of the slope corresponds to the sign of the transition frequency listed
in Table~\ref{tab:transitions}. The expected high-detuning positions of the smaller peaks are indicated by dashed lines. Additional weaker features can be attributed to degenerate two-photon RF resonances for $70S_{1/2}\to69S_{1/2}$ and $70S_{1/2}\to71S_{1/2}$ transitions.  Here two microwave photons are absorbed simultaneously and therefore excite a transition with double the frequency of the photons. Changing the frequency changes the frequency of both photons, hence $\delta_\mathrm{RF} = 2f_\mathrm{RF}-f_0$. This produces branches with slopes of $\pm 1/2$ in respect to $\Delta_c$. Their expected positions are indicated by the red lines in Fig.~\ref{fig:fig3}.

At the lower RF power shown in Fig.~\ref{fig:fig3}(b), the RF coupling is weaker and the observed branches more closely follow the large-detuning behavior of Eq.~\eqref{eq:detuning}. Additionally, the degenerate two-photon transitions are invisible since it is a second-order process.

In contrast to Fig.~\ref{fig:fig3}(b), Fig.~\ref{fig:fig3}(c) was measured at a
significantly higher RF power of $8~\mathrm{dBm}$. At this power, the
RF-induced splitting is dominated by the Rabi-frequency contribution in
Eq.~\eqref{eq:coupling}. In addition, the interaction with
multiple nearby Rydberg levels becomes more pronounced, particularly in the
frequency range between the $70P_{3/2}$ and $69P_{3/2}$ transitions, where additional spectral structure is observed.

\subsection{Frequency--Power Response}
\label{subsec:response}

\begin{figure}[h]
  \centering
  \includegraphics{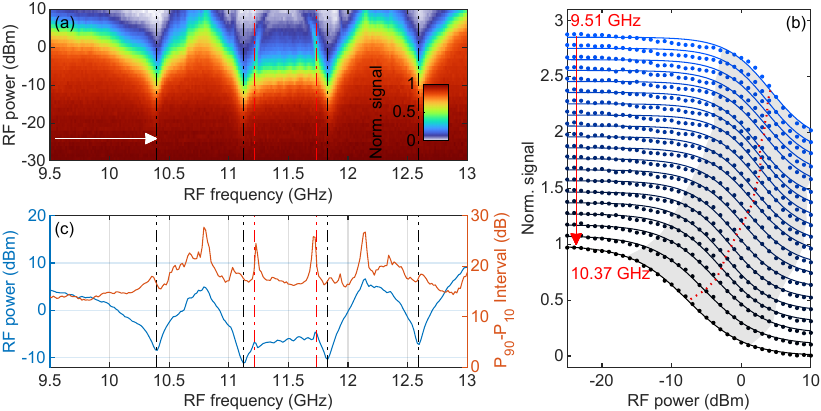}
  \caption{(a) False color plot of EIT transmission signal at zero control field detuning for different RF powers and frequencies. Four vertical lines mark the theoretical values of transition frequencies. Black lines correspond to single-photon transitions and the red lines correspond to degenerate two-photon transitions. (b) Stacked plot depicting EIT transmission signal at $\Delta_c=0$ for a range of frequencies below and up to  transition. Plot range is marked using a white arrow in (a). Shaded area depicts the range where the signal goes from 90\% to 10\%. (c) Plot of RF powers where EIT transmission signal falls to a half of its total variation (blue) and the interval where the transmission signal falls from 90\% to 10\% (orange).}
  \label{fig:fig4}
\end{figure}

Fig.~\ref{fig:fig4}(a) shows the transmission signal as a function of
RF frequency and applied RF power for zero coupling-laser detuning. Spectral
features associated with the same four Rydberg--Rydberg transitions can again
be observed, together with the aforementioned degenerate two-photon RF resonance. The
white arrow marks the RF frequencies at which vertical cuts through the map
were taken, with the corresponding stacked line profiles shown in
Fig.~\ref{fig:fig4}(b). We characterize the power-dependent response using a sigmoid function
\begin{equation}
S(P)=
\frac{S_{0}}
{1+\exp[-(P-P_{0})/w]},
\label{eq:sigmoid}
\end{equation}
where $S_0$ is the fitted signal amplitude, $P_0$ is the midpoint of the
transition, and $w$ characterizes its width. Both the position and the width of
the transition depend on the RF frequency. We have chosen the sigmoid function since the complex susceptibility of a three-level system at zero-detuning has the form of a sigmoid function when using decibel units for coupling intensity \cite{Lambropoulos2007}.

Figure~\ref{fig:fig4}(c) shows the fit midpoint $P_0$ and the transition
width $P_{90}-P_{10}$ as functions of RF frequency. Here, $P_{10}$ and
$P_{90}$ correspond to RF powers at which the fitted signal reaches $10\%$
and $90\%$ of its total variation, respectively. This interval defines the
range of RF powers over which the system exhibits an approximately logarithmic
RF-to-optical response. The lowest values of $P_0$, and therefore the highest
sensitivity, occur at the Rydberg--Rydberg transition frequencies. In
contrast, the degenerate two-photon resonances exhibit a larger $P_{90}-P_{10}$ interval,
corresponding to a larger dynamic range at the expense of somewhat lower
sensitivity.

\section{Conclusion}
In summary, we provide a detailed measurement and analysis of the optical response of a room-temperature cesium Rydberg microwave receiver for a broad range
of RF frequencies and powers. We clearly resolve four Rydberg-Rydberg transitions and identify a pair of weaker degenerate two-photon Rydberg-Rydberg transitions. The broad measurement range allows us to observe both the large-detuning limit and the strong-coupling regime. By fitting the measured spectra we are able to determine the $70S_{1/2}$ to $69P_{1/2}$, $69P_{3/2}$, $70P_{1/2}$ and $70P_{3/2}$ transition frequencies.

We further analyze the power-dependent optical signal and identify a range in which the system can operate as a logarithmic RF-to-optical receiver. As expected, we confirm the highest responsivity near Rydberg-Rydberg transitions. Additionally, we observe a relatively uniform responsivity range with approximately 15 dB of dynamic range. Lastly, we observe that in proximity to the degenerate two-photon transition the dynamic range expands to about 25 dB with slightly lower responsivity when compared to single-photon transitions. We believe that this is an accessible trade-off for applications that prioritize dynamic range over peak responsivity.

The same experimental and analysis approach can be extended to other Rydberg states and alkali-metal vapor-cell systems, allowing the RF-frequency and response of the receiver to be selected for a specific application by choice of the coupled Rydberg pair. More broadly, the frequency- and power-resolved characterization presented here provides a general framework for analyzing Rydberg-EIT receivers for their use in RF sensing, metrology, and communication applications.

\bibliographystyle{IEEEtran}
\bibliography{references} 

\end{document}